\documentclass[reprint,
superscriptaddress,amsmath,amssymb,aps,prl]{revtex4-2}
\usepackage{graphicx}
\usepackage{dcolumn}
\usepackage{bm}
\usepackage{xcolor}
\usepackage[normalem]{ulem}
\usepackage{CJK}
\usepackage{times, mathptmx}
\usepackage{anyfontsize}

\begin{document}
\begin{CJK*}{UTF8}{bsmi}


\title{\textbf{Electronic reconstruction at the quasicrystal--moir\'{e} crossover in twisted bilayer graphene} 
}

\author{Kuo-En Chang}
\altaffiliation{These authors contributed equally to this work.}
\affiliation{Department of Physics, National Cheng Kung University, Tainan 701, Taiwan}
\affiliation{Center for Quantum Frontiers of Research \& Technology (QFort), National Cheng Kung University, Tainan 701, Taiwan}

\author{Aitor Garcia-Ruiz (艾飛宇)}
\altaffiliation{These authors contributed equally to this work.}
\affiliation{Department of Physics, National Cheng Kung University, Tainan 701, Taiwan}
\affiliation{Center for Quantum Frontiers of Research \& Technology (QFort), National Cheng Kung University, Tainan 701, Taiwan}

\author{Ta-Lei Chou}
\affiliation{Center for Condensed Matter Sciences, National Taiwan University, Taipei 106, Taiwan}
\affiliation{Center of Atomic Initiative for New Materials, National Taiwan University, Taipei 106, Taiwan}

\author{Yen-Ting Liu}
\affiliation{Department of Physics, National Cheng Kung University, Tainan 701, Taiwan}

\author{Sheng-Chin Ho}
\affiliation{Department of Physics, National Cheng Kung University, Tainan 701, Taiwan}

\author{Yu-Chiang Hsieh}
\affiliation{Department of Physics, National Cheng Kung University, Tainan 701, Taiwan}
\affiliation{Center for Quantum Frontiers of Research \& Technology (QFort), National Cheng Kung University, Tainan 701, Taiwan}

\author{Ching-Hua Kao (高慶樺)}
\affiliation{Center for Quantum Frontiers of Research \& Technology (QFort), National Cheng Kung University, Tainan 701, Taiwan}

\author{Chiu-Hua Huang}
\affiliation{Department of Physics, National Cheng Kung University, Tainan 701, Taiwan}

\author{Ying-Mei Yang}
\affiliation{Center for Quantum Frontiers of Research \& Technology (QFort), National Cheng Kung University, Tainan 701, Taiwan}

\author{Kenji Watanabe}
\affiliation{Research Center for Electronic and Optical Materials, National Institute for Materials Science, Namiki 1-1, Tsukuba, 305-0044, Ibaraki, Japan}

\author{Takashi Taniguchi}
\affiliation{Research Center for Materials Nanoarchitectonics, National Institute for Materials Science, Namiki 1-1, Tsukuba, 305-0044, Ibaraki, Japan}

\author{Ming-Wen Chu}
\affiliation{Center for Condensed Matter Sciences, National Taiwan University, Taipei 106, Taiwan}
\affiliation{Center of Atomic Initiative for New Materials, National Taiwan University, Taipei 106, Taiwan}

\author{Ming-Hao Liu (劉明豪)}
\email{minghao.liu@phys.ncku.edu.tw}
\affiliation{Department of Physics, National Cheng Kung University, Tainan 701, Taiwan}
\affiliation{Center for Quantum Frontiers of Research \& Technology (QFort), National Cheng Kung University, Tainan 701, Taiwan}

\author{Tse-Ming Chen}
\email{tmchen@phys.ncku.edu.tw}
\affiliation{Department of Physics, National Cheng Kung University, Tainan 701, Taiwan}
\affiliation{Center for Quantum Frontiers of Research \& Technology (QFort), National Cheng Kung University, Tainan 701, Taiwan}

\begin{abstract}

Large twist angles in twisted bilayer graphene are widely expected to be electronically trivial, with negligible interlayer coupling and no electronic reconstruction, in contrast to the rich moir\'{e}-driven band reconstruction and correlated physics that emerge at small twist angles. Here, we show that this paradigm breaks down near a twist angle of $29^\circ$, where the system crosses over between quasicrystalline and commensurate order. Atomic-resolution transmission electron microscopy directly reveals the coexistence of near-dodecagonal quasicrystalline symmetry and emerging moir\'{e} periodicity, indicating an intermediate, nonperiodic structural regime. Magnetotransport measurements uncover strong interlayer hybridization mediated by Umklapp scattering, manifested by magneto-intersubband oscillations and a highly unconventional Landau-level spectrum. Remarkably, the Landau-level degeneracy evolves from fourfold to twelvefold with increasing temperature, a behavior incompatible with two decoupled graphene monolayers. These findings establish large-angle twisted bilayer graphene as a platform where quasiperiodic symmetry fundamentally reshapes low-energy electronic states beyond the conventional moir\'{e} framework.
\end{abstract}

\maketitle
\end{CJK*}
Twistronics has emerged as a powerful platform for engineering lattice geometry and electronic structure in van der Waals (vdW) heterostructures, where the relative orientation between atomically thin layers provides an exceptional degree of control over interlayer coupling and band topology~\cite{bistritzer2011moire, andrei2020graphene, yoo2019atomic}. A paradigmatic example is small-angle twisted bilayer graphene (TBG), in which a twist near the magic angle generates long-wavelength moir\'{e} superlattices and flat bands that host strongly correlated electronic phases~\cite{cao2018correlated, cao2018unconventional, yankowitz2019tuning, sharpe2019emergent, uri2023superconductivity}. In contrast, the large-angle regime has received comparatively little attention. The prevailing expectation has been that short-wavelength lattice mismatch at large twist angles leads to band reconstructions far from the Fermi level, rendering the system effectively equivalent to two decoupled layers---an assumption supported by early transport measurements at large angle~\cite{rickhaus2020electronic, pezzini202030, mrenca2022quantum, babich2025milli}.

This picture has recently been challenged by experiments on TBG and WSe$_2$ precisely at a twist angle of $30^\circ$, where the system forms an incommensurate dodecagonal quasicrystal~\cite{uri2023superconductivity,liu2025field}. These studies revealed anomalously strong interlayer coupling mediated by Umklapp scattering, giving rise to multiple Dirac-cone replicas and a quasiperiodic, rotational-symmetry-driven reconstruction of the electronic structure~\cite{ahn2018dirac, yao2018quasicrystalline, moon2019quasicrystalline, li2024tuning}. The emergence of vdW quasicrystals thus opens a new avenue for exploring coupling mechanisms beyond conventional moir\'{e} descriptions, as well as exotic topological states and correlated quantum phases enabled by quasiperiodic order~\cite{cao2020kohn, else2021quantum, koshino2022topological, liu2023high, tsang2024polar, ghadimi2025quasiperiodic}. Despite these advances, the strong interlayer hybridization observed to date is predominantly limited to high-energy regimes, leaving the system's low-energy transport behavior indistinguishable from that of decoupled layers~\cite{babich2025milli, pezzini202030}. Moreover, the crossover from this aperiodic quasicrystalline order to the translational invariance of commensurate moir\'{e} superlattices remains largely unexplored. Investigating this transition by continuously tuning the twist angle away from the $30^\circ$ configuration provides a unique opportunity to elucidate how competing symmetries and superlattice periodicity govern low-energy electronic properties.

\begin{figure*}
\centering
    \includegraphics[width=0.7\textwidth]{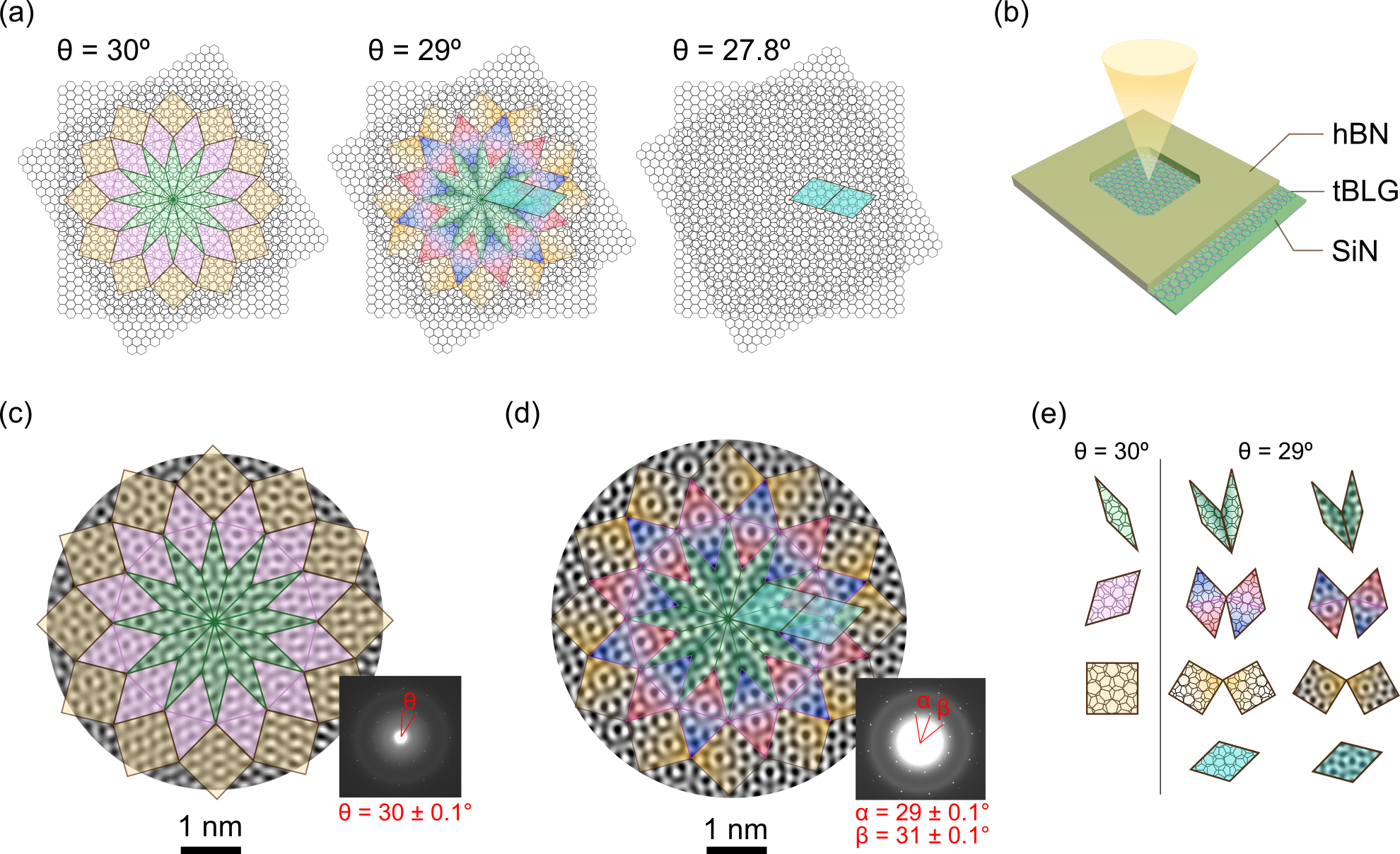}
\caption{\label{fig:1} (a) Schematic illustrations of atomic arrangements from the $30^\circ$ quasicrystal (left), with 12-fold dodecagonal symmetry described by a Stampfli tiling, to the $27.8^\circ$ commensurate moir\'{e} superlattice (right). At $\theta = 29^\circ$, the quasicrystalline order is gently perturbed, producing a crossover state that retains mirror-symmetric motifs of the dodecagonal structure while showing incipient local moir\'{e} periodicity near the rotating center. (b) Schematic of the device structure incorporating a prepatterned access-window in the hBN pickup layer for atomic-resolution TEM imaging. (c) and (d) show the TEM image of $30^\circ$ and $29^\circ$ TBG, respectively. In (c), the image reveals a dodecagonal quasicrystalline lattice consistent with a Stampfli tiling. In contrast, (d) shows a perturbation of the dodecagonal order, reduced to 6-fold rotational symmetry with mirror-symmetric motifs. The insets display the corresponding diffraction patterns, which are directly measured from the selected-area electron diffraction (SAED). (e) Direct comparison of color-coded tilings at $30^\circ$ and $29^\circ$, with gradient shading highlighting spatial variations and mirror-related motifs; local moir\'{e} periodicity is also visible in the $29^\circ$ sample.}
\end{figure*}

In this paper, we investigate the low-energy electronic properties of TBG at a twist angle of $29^\circ$, lying just $1^\circ$ away from the dodecagonal quasicrystalline phase at $30^\circ$. Combining large-scale transport simulations with magnetotransport measurements on a dual-gated device, we uncover clear signatures of interlayer Umklapp scattering that strongly influence the low-energy electronic response, even in the absence of conventional moir\'{e} band reconstruction. Magnetic oscillations in the longitudinal resistance reveal anomalously strong interlayer coupling. More strikingly, the Hall conductance exhibits an unconventional temperature-dependent evolution of Landau-level degeneracy, transitioning from fourfold to twelvefold. These observations are incompatible with a description based on two decoupled graphene monolayers and establish $29^\circ$ TBG as a model system for exploring the crossover between quasicrystalline and translationally invariant regimes.

Figure~\ref{fig:1}(a) illustrates the structural evolution of TBG near $\theta = 30^\circ$. At $\theta = 30^\circ$ (leftmost), TBG forms a dodecagonal quasicrystal with 12-fold rotational symmetry and no translational periodicity, well described by a Stampfli tiling~\cite{stampfli1986dodecagonal, ahn2018dirac}. A small deviation from this angle, as in $29^\circ$-TBG (middle panel), perturbs the tiling and reduces the symmetry to sixfold. This yields mirror-symmetric motifs that, despite lacking long-range translational invariance, exhibit a short-range resemblance to the dodecagonal quasicrystal and geometrically interpolate towards the nearby commensurate moir\'{e} structure. With further deviation from $30^\circ$, the structure evolves towards a commensurate moir\'{e} superlattice, with $\theta = 27.8^\circ$ (rightmost) representing the nearest low-order commensurate phase. The $29^\circ$ configuration thus constitutes an intermediate geometric regime, combining elements of quasicrystalline and emerging moir\'{e} order and providing a natural setting to explore the crossover between rotational and translational symmetries (Sec.~3 in SM ~\cite{SI}).

This intermediate state can be directly visualized using transmission electron microscopy (TEM) on our TBG samples. Most TEM investigations of TBG have so far focused on CVD or epitaxial bilayers, which can be prepared as large-area freestanding membranes compatible with standard TEM holders~\cite{ahn2018dirac, tsang2024polar}. In such samples, however, the twist angle is growth-determined and cannot be arbitrarily and precisely tuned. More recently, TEM has also been applied to tear-and-stack TBG, which enables investigations into magic-angle regions~\cite{yoo2019atomic, kazmierczak2021strain, li2025robust}. However, the thick hBN pickup flake essential to this stacking process can attenuate lattice contrast and limit the achievable resolution; consequently, these studies have typically emphasized diffraction, dark-field, or four dimensional scanning transmission electron microscopy (4D-STEM) mapping of moir\'{e} order rather than direct atomic-resolution imaging. Here, we overcome this limitation by using an hBN pickup flake with a prepatterned access-window, fabricated by electron-beam lithography and dry etching (Sec.~1 in SM~\cite{SI}), to perform the tear-and-stack assembly and prepare TBG TEM samples, as shown in Fig.~\ref{fig:1}(b).

Figure~\ref{fig:1}(c) shows an atomic-resolution TEM image of $30^\circ$ TBG. The inset, a selected area electron diffraction (SAED) pattern taken from this region, confirms a precisely aligned quasicrystalline structure with dodecagonal symmetry, consistent with previous reports.~\cite{ahn2018dirac, vidarte2024quasicrystalline, stampfli1986dodecagonal}. Corresponding measurements on the $29^\circ$-TBG sample [Figs.~\ref{fig:1}(d)] reveal a clear departure from quasicrystalline order, characterized by reduced sixfold rotational symmetry and pronounced local lattice variations. Notably, enhanced moir\'{e}-like contrast is observed in certain regions, bearing visual resemblance to the nearby commensurate phase at $\theta = 27.8^\circ$ and appearing significantly stronger than expected from a rigid-lattice model. This enhancement may arise from local strain or atomic reconstruction~\cite{yoo2019atomic, kazmierczak2021strain}. A direct comparison between the color-coded tilings at $\theta = 30^{\circ}$ and $29^{\circ}$ is shown in Fig.~\ref{fig:1}(e), where gradient shading is used to mark the spatial lattice variations and to help visualize the mirror symmetry. In contrast to these prominent local motifs, large-area TEM [Fig.~S2] shows no indication of long-range commensurate order, suggesting that such periodic features are strictly localized. The broader macroscopic properties are discussed in SM Sec.~2 and Sec.~3~\cite{SI}.

\begin{figure*}
\centering
    \includegraphics[width=0.8\textwidth]{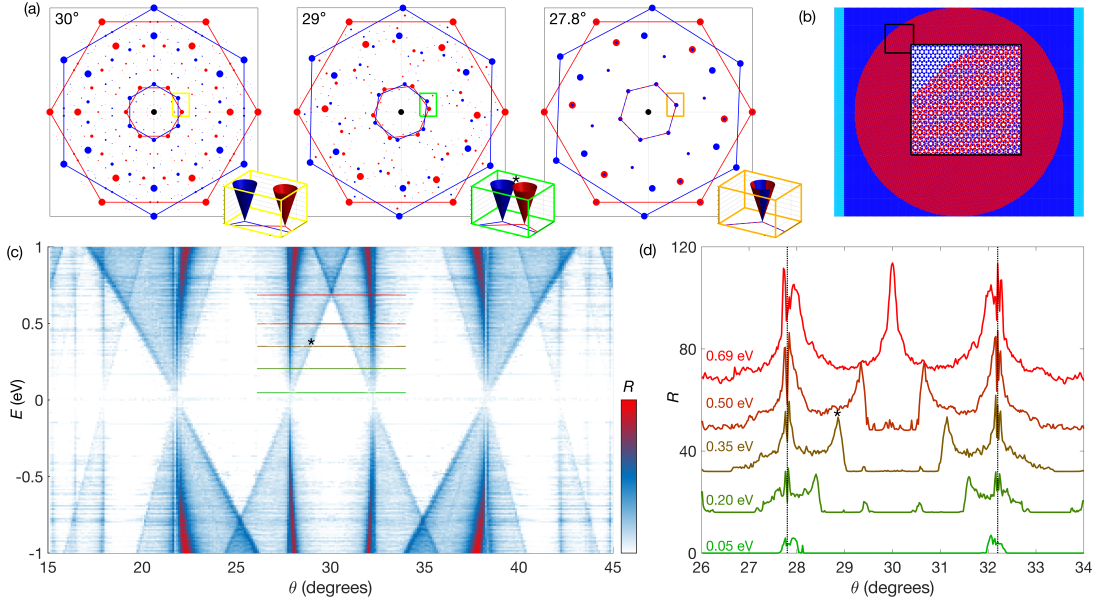}
\caption{(a) Sketch presenting the corners of the first Brillouin zone and the Umklapp-scattered points $\mathbf{k}=\mathbf{K}_{t/b}^{(\pm)}+\mathbf{G}_\mathrm{b}-\mathbf{G}_\mathrm{t}$. The size of the points reflects the amplitude of the spectral weight, exponentially decaying with $\mathbf{G}_\mathrm{b}-\mathbf{G}_\mathrm{t}$. The inset presents the low-energy decoupled dispersion, from the charge neutrality point to $E_{\mathrm{F}}=0.4$ eV, around the second-order Umklapp-scattered dispersion. (b) Sketch of the device used in the transport simulations. The blue and red dots represent the atoms of the bottom and top flakes, and the cyan atoms represent the leads. (c) Reflection map for a 50 nm-wide device. (d) Reflection profile for 5 selected Fermi levels, for a 100 nm-wide nanoribbon, with the traces vertically offset for clarity. To note, the strength of the Umklapp-assisted peak is comparable to the commensurate-moir\'{e} peak.\label{fig:2}}
\end{figure*}

The relative twist between the two graphene layers directly determines the structure of the electronic states in reciprocal space, manifesting as a corresponding rotation of their Brillouin zones. At $\theta = 30^\circ$, the $K$ points of the two layers form a perfect dodecagonal arrangement, as illustrated in Fig.~\ref{fig:2}(a). Interlayer coupling allows electrons to scatter between inequivalent valleys via reciprocal lattice vectors of the two layers, satisfying the interlayer Umklapp condition $
\mathbf{k}+\mathbf{G}_\mathrm{t}=\mathbf{k}'+\mathbf{G}_\mathrm{b}$.
This process gives rise to multiple Dirac-cone replicas, as directly observed by ARPES in $30^\circ$ TBG~\cite{ahn2018dirac}. These replicas can weakly hybridize, providing additional channels for interlayer coupling. As the twist angle is reduced from $30^\circ$, the Umklapp-induced Dirac states of the two layers hybridize at progressively lower energies, approaching the Fermi level. At the nearby commensurate angle $\theta = 27.8^\circ$, Umklapp states from the two layers share a common momentum origin, marking the emergence of translationally invariant moir\'{e} electronic structure.

To investigate the crossover regime, we simulate charge transport through a two-terminal TBG device, schematically shown in Fig.~\ref{fig:2}(b), following the real-space Green's functions method~\cite{datta_electronic_1995, chakraborti2024electron}. The system consists of a 50-nm-wide zigzag graphene nanoribbon connected to semi-infinite graphene leads, with a circular graphene flake twisted by an angle $\theta$ placed in the central scattering region. Interatomic couplings are modeled using a Slater--Koster parametrization (Sec.~4 in SM~\cite{SI}). In this geometry, weak interlayer coupling leaves transport dominated by the nanoribbon, whereas enhanced interlayer hybridization causes the twisted flake to act as an effective backscattering center. The energy- and angle-dependent reflection thus provides a direct probe of the interlayer coupling.

Figure~\ref{fig:2}(c) shows the calculated electron reflection as a function of twist angle $\theta \in~[15^\circ,45^\circ]$ and Fermi energy within $\pm1$ eV relative to the charge neutrality. Two distinct families of high-reflection features emerge. The first consists of vertical stripes at specific commensurate angles, reflecting enhanced interlayer coherence arising from geometric commensurability. As a purely structural effect, these features are nearly energy-independent. The second family exhibits diagonally dispersing features that originate from interlayer Umklapp scattering.

\begin{figure*}
\centering
    \includegraphics[width=0.7\textwidth]{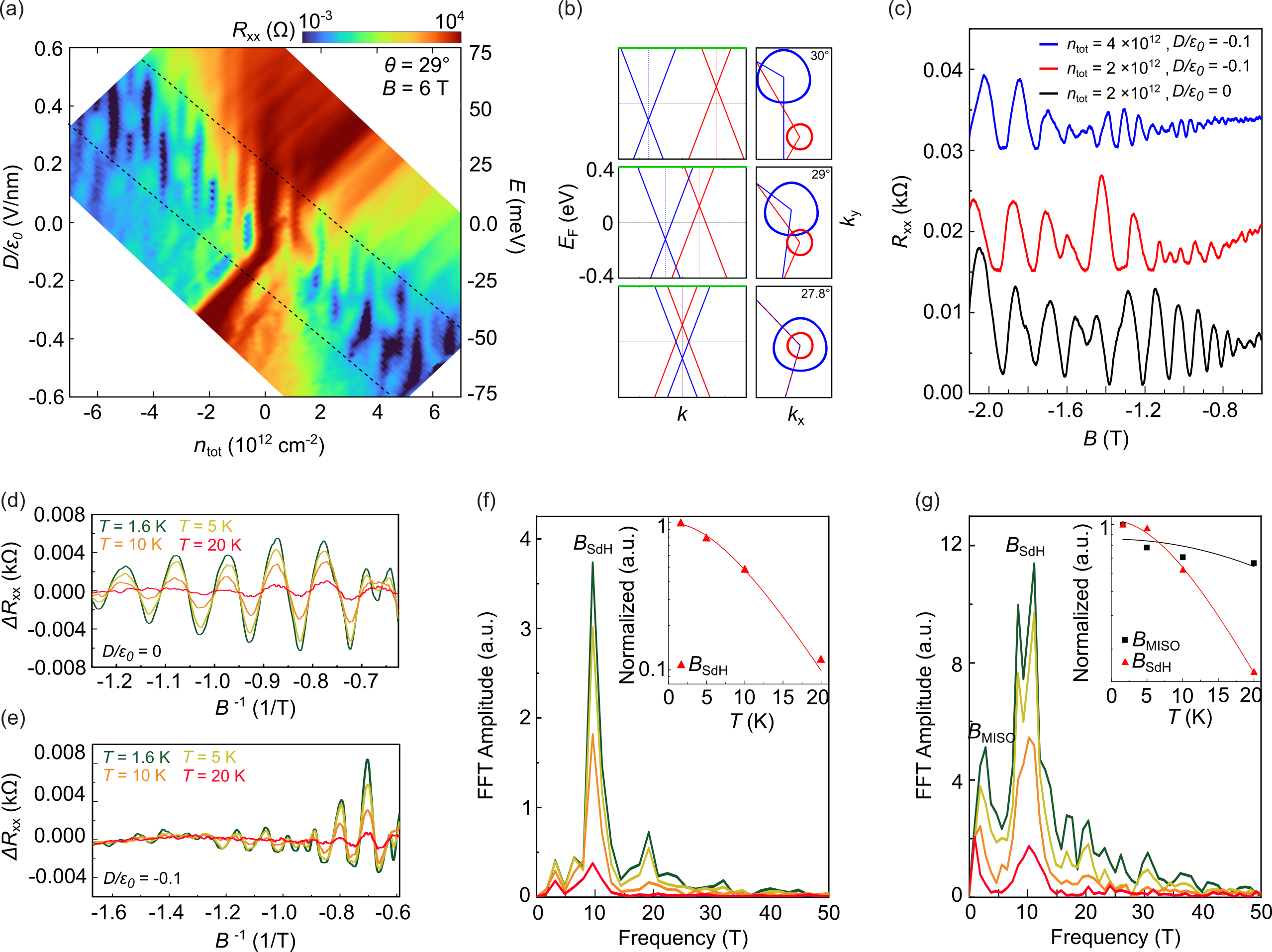}
\caption{\label{fig:3}(a) Longitudinal resistance $R_{\mathrm{xx}}$ of dual-gated $29^\circ$ TBG (device $B_{29}$) as a function of carrier density $n_{\mathrm{tot}}$ and displacement field $D$ at $B = 6$ T and $T = 1.6$ K, revealing a distorted and asymmetric Landau level spectrum. (b) Calculated Fermi contours of second-order Umklapp Dirac-cone replicas at three twist angles under a displacement field ($B$ = 0~T), obtained from a line cut at $E_{\mathrm{F}}$ = 0.4~eV (green trace in the left panel). (c) Low-field longitudinal resistance $R_{\mathrm{xx}}$ as a function of magnetic field $B$ for different displacement fields~$D$ and carrier densities $n_{\mathrm{tot}}$. (d) and (e) Temperature dependence of oscillations plotted against inverse magnetic field ($1/B$) at $D/\epsilon_0 = 0$ and $D/\epsilon_0 = -0.1$~V/nm, respectively, both at $n_{\mathrm{tot}} = 2 \times 10^{12}$ cm$^{-2}$. (f) and (g) Corresponding fast Fourier transform (FFT) spectra of (d) and (e). Peaks labeled $B_{\mathrm{SdH}}$ correspond to conventional SdH oscillations and quantify the carrier density, while the additional $B_{\mathrm{MISO}}$ component reflects the magneto-intersubband oscillations arising from interlayer scattering. Insets: temperature dependence of normalized FFT amplitudes. The SdH peaks follow the Lifshitz--Kosevich relation, whereas the MISO component decays more slowly with temperature, consistent with its scattering-driven origin.}
\end{figure*}

To determine which of these features could be relevant in low-energy transport experiments, we analyze their energy evolution in Fig.~\ref{fig:2}(d), which tracks the reflection peaks as a function of twist angle at fixed energies. At $\theta=30^\circ$, the Umklapp-induced features occur at relatively high energies ($\sim 0.7$ eV), and remain above $\sim 0.5$ eV for angles close to $29.4^\circ$, making them difficult to access experimentally. As the twist angle is reduced, these features shift toward lower energies, reaching $\sim 0.35$ eV at $\theta=29^\circ$, where they fall within an experimentally relevant energy window. For angles closer to the commensurate value $\theta_{1,3}=27.8^\circ$, the characteristic energy scale is further reduced; however, in this regime the system becomes increasingly dominated by commensurate order, and the corresponding reflection features decrease in intensity. 

We therefore identify $\theta \approx 29^\circ$ as an optimal regime in which strong Umklapp-induced scattering occurs at low but finite energies, while the system remains distinct from the fully commensurate limit. This balance makes it a particularly suitable platform to probe the interplay between quasicrystalline and commensurate physics in the low-energy transport regime.

To probe the impact of interlayer Umklapp scattering on low-energy transport, we fabricated encapsulated dual-gated TBG Hall bar devices with twist angles $\theta = 30^\circ$, $29^\circ$, and $27.8^\circ$ using a tear-and-stack dry transfer technique (Sec.~1 in SM~\cite{SI}). Independent control of the total carrier density $n_{\mathrm{tot}}$ and perpendicular displacement field $D$ allows us to map the longitudinal resistance $R_\mathrm{xx}(n_{\mathrm{tot}},D)$ at finite magnetic field, providing direct access to the Landau-level structure. For effectively decoupled graphene monolayers, this map exhibits characteristic diamond-shaped patterns arising from Landau-level crossings~\cite{slizovskiy_out--plane_2021,sanchez2012quantum, kim2021odd, tomic2022scattering}, which we indeed observe in devices with $\theta = 30^\circ$ and $27.8^\circ$  (Sec.~6 in SM~\cite{SI}). In striking contrast, the $\theta = 29^\circ$ device displays pronounced qualitative deviations as shown in Fig.~\ref{fig:3}(a), including electron--hole asymmetry and inversion-symmetry breaking with respect to $D$. The corresponding energy scale associated with the applied displacement field is on the order of $\Delta E \sim e D d / \varepsilon_{\mathrm{GG}}$ with $d \sim 0.335$ nm and $\varepsilon_{\mathrm{GG}} \sim 2.5$~\cite{slizovskiy_out--plane_2021, sanchez2012quantum}. The distorted diamond-like structure is only discernible within a narrow, diagonally oriented stripe in the ($n_\mathrm{tot},D$) plane, suggesting a correlated constraint between displacement field and carrier density that is absent in decoupled monolayers. Such behavior can be attributed to magnetic breakdown, a form of band entanglement in which tunneling between nearby, layer-polarised Fermi contours become allowed, as schematically illustrated in Fig.~\ref{fig:3}(b). Similar distorted $n_\mathrm{tot}$--$D$ transport features are reproduced in additional $29^\circ$ devices, as shown in the SM~\cite{SI}.

\begin{figure*}
\centering
    \includegraphics[width=0.8\textwidth]{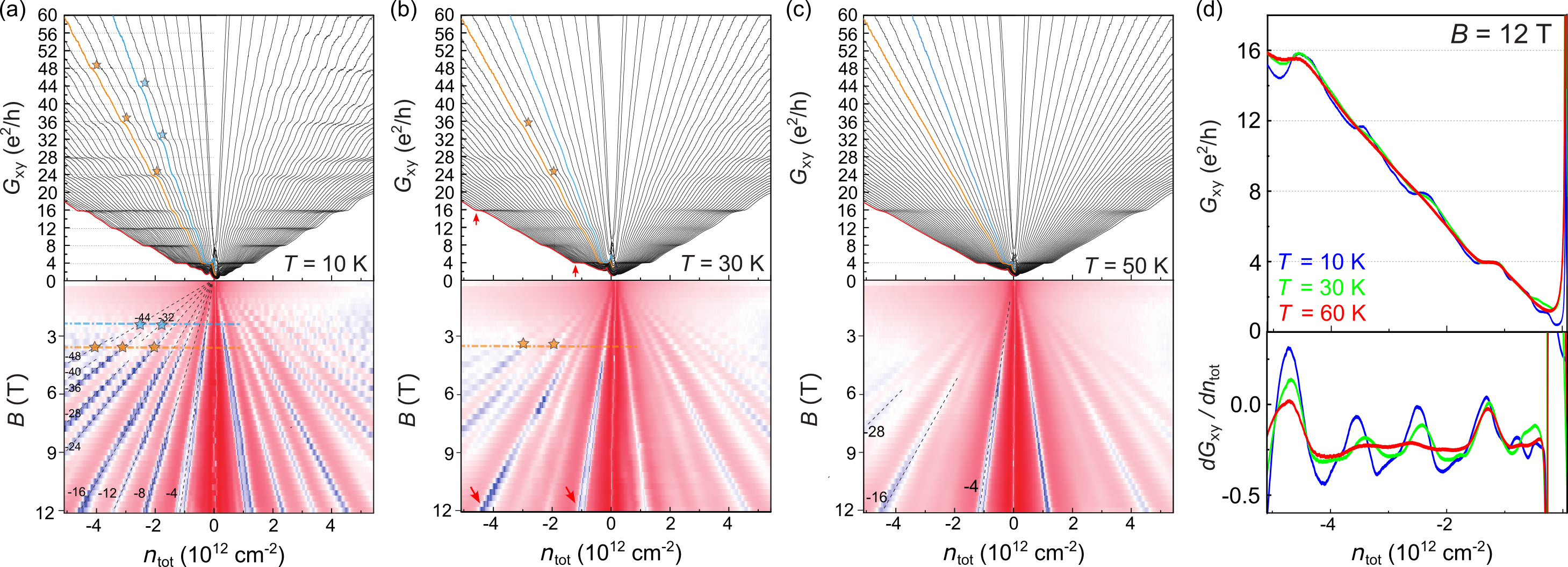}
\caption{\label{fig:4}(a)-(c) Hall conductance $G_{\mathrm{xy}}$ (upper panels) and longitudinal resistance $R_{\mathrm{xx}}$ (lower panels) as functions of carrier density and magnetic field at $T = 10, 30, 50$~K respectively. Star markers highlight the quantum Hall plateaus associated with the $12e^2/h$ sequence in selected regimes. Dashed lines trace the corresponding Landau levels, making the sequence easier to follow. In (b), arrows indicate the comparatively robust $\nu = 4$ and $16$ plateaus. (d) Temperature dependence of $G_{\mathrm{xy}}$ (top) at $B = 12$~T and its derivative (bottom), illustrating the anomalous robustness of the $4e^2/h$ and $16e^2/h$ plateaus.}
\end{figure*}

To independently examine whether interlayer coupling is present in this regime, we measured Shubnikov--de Haas (SdH) oscillations at finite doping [Fig.~\ref{fig:3}(c)]. For zero interlayer bias [Fig.~\ref{fig:3}(d, f)], the frequency spectrum is consistent with two identical circular Fermi surfaces ($E_\mathrm{F}=\hbar v\sqrt{\pi\vert n \vert/2}\approx 0.1$ eV). The dominant frequency $B_\mathrm{SdH} \approx 10$~T extracted from the $1/B$ analysis yields a carrier density consistent with our capacitance model. When a displacement-field bias is applied [Fig.~\ref{fig:3}(e, g)], two closely spaced SdH frequencies at $\sim9$ T and $\sim11$ T emerge, together with an additional low-frequency peak around $\sim2$~T. This low-frequency oscillation arises from the beating of the two higher frequencies and is a hallmark of magneto-intersubband oscillations (MISO), originating from magnetic-field-assisted scattering between distinct Fermi contours~\cite{fallahazad_quantum_2012,nitta1997gate,lo2017controlled,phinney2021strong,yuan2024interplay}. Since these contours in TBG are predominantly localized in opposite layers, the appearance of MISO directly confirms finite interlayer coupling and magnetically entangled subbands~\cite{sanchez2012quantum, phinney2021strong,yuan2024interplay}. For a detailed density-dependent analysis of this intersubband interaction, see Sec.~8 in SM~\cite{SI}. Consistently, the MISO peak exhibits a markedly different temperature dependence from SdH oscillations, while SdH amplitudes follow the Lifshitz--Kosevich form, the MISO signal decays more slowly, approximately as $\exp(-\gamma T^2)$, with $\gamma\approx8.3\times10^{-4}$~K$^{-2}$ in our data (Sec.~8 in SM~\cite{SI}). Consequently, MISO remains prominent up to $\sim20$~K [Fig.~\ref{fig:3}(g)], demonstrating that interlayer coupling persists well beyond the thermal smearing limit of conventional SdH oscillations in small-angle TBG. To note, this temperature also provides a lower bound for the interlayer coupling of 2~meV.

Further evidence for finite interlayer coupling is provided by the Landau level degeneracy and its temperature dependence in devices with twist angles $\theta=30^\circ,29^\circ$ and $27.8^\circ$. The expected eightfold (spin, valley and layer) sequence of electronically decoupled monolayers is indeed observed in both the $30^\circ$ and $27.8^\circ$ samples. In stark contrast, the $29^\circ$ device exhibits a fourfold degeneracy sequence at low temperatures [Fig.~\ref{fig:4}(a)], indicating that even near the charge neutrality point the electronic spectrum no longer resembles that of two independent monolayers. While interlayer Umklapp-assisted magnetic breakdown could, in principle, reduce the effective degeneracy by enabling tunneling between nearby layer-resolved orbits, calculations of the Landau level spectrum for commensurate TBG nanoribbons do not fully account for the observed fourfold sequence~\cite{SI}, pointing to a more intricate reconstruction of the low-energy states in this crossover regime.

Even more strikingly, upon increasing the temperature, a twelvefold quantum Hall sequence emerges, a phenomenon that has not been previously reported in graphene-based systems. Figures~\ref{fig:4}(b) and (c) show the Hall conductance and the corresponding Landau fan diagrams at $T=30$ and $50$ K. Notably, the quantum Hall features exhibit a nonuniform temperature evolution. While most plateaus and corresponding longitudinal resistance minima in the Landau fan are progressively smeared out upon warming, those associated with the dodecagonal-like degeneracy remain relatively robust. At $T = 50$~K, well-defined quantum Hall signatures persist at filling factors $\nu = 4$, $16$, and $28$. The evolution from a fourfold to a twelvefold degeneracy is further illustrated in Fig.~\ref{fig:4}(d), which compares the Hall conductance and its derivative at a fixed magnetic field across different temperatures. This nontrivial temperature-driven transition suggests the activation of additional low-energy states associated with interlayer Umklapp processes, and we identify it as a defining transport signature of the quasicrystal-moir\'{e} crossover at $\theta=29^\circ$ (see Fig.~S15 for reproducibility). We also note that signatures of a twelvefold quantum Hall sequence can be discernible at low temperature and higher filling factors; for example, at $B = 3.6$~T, plateaus appear at $24e^2/h$, $36e^2/h$, and $48e^2/h$. Several theoretical studies of near-commensurate TBG have predicted the emergence of flat bands~\cite{pal_theory_2014,pal_emergent_2019}, which can amplify electron-electron interaction effects. While such flat-band scenarios alone do not explain the observed anomalous temperature dependence, they point toward an intriguing regime, where interaction effects may intertwine with Umklapp-induced hybridization. This remains as an important direction for future work.

Our findings, including the observation of pronounced interlayer coupling and dodecagonal Landau degeneracy, raise intriguing questions about the nature of the electronic structure in this $29^\circ$ quasicrystal--moir\'{e} crossover. Anomalously strong interlayer couplings have been consistently reported in $30^\circ$ TBG quasicrystal~\cite{ahn2018dirac}. This significantly enhances higher-order Umklapp scattering processes and gives rise to 12-fold symmetric Dirac cone replicas near the $\Gamma$, whose experimentally observed intensities exceed the theoretical predictions by several orders of magnitude~\cite{ahn2018dirac}. These replicas provide an extended framework for inter-replica hybridization, potentially leading to the formation of additional minibands at lower energies. The thermally assisted emergence of dodecagonal Landau-level degeneracy in the low-energy transport regime further suggests that, in this near-quasicrystalline regime, phonon- and phason-mediated tunneling may facilitate interlayer coupling pathways, enabling miniband formation from higher-order Umklapp-scattered Dirac cone replicas~\cite{ochoa2019moire, samajdar2022moire, ochoa2022degradation}. Moreover, the incipient formation of moir\'{e} periodicity observed structurally at $29^\circ$ raises the possibility of minibands and layer hybridization through moir\'{e}-induced scattering~\cite{hamer2022moire}, albeit modulated by residual quasiperiodic symmetry. The coexistence and interplay of these two mechanisms may give rise to an unconventional miniband structure, with hybrid characteristics that reflect both rotational and translational symmetry breaking.

In conclusion, we demonstrate that large-angle TBG is not generically decoupled, near $29^\circ$, quasiperiodic symmetry can drive pronounced low-energy electronic reconstruction. This structure, bridging the incommensurate quasicrystalline phase at $30^\circ$ and the commensurate moir\'{e} superlattice at $27.8^\circ$, reveals a unique crossover between rotational and translational symmetry, directly visualized through atomic-resolution transmission electron microscopy. Despite the absence of conventional moir\'{e} band reconstruction, magnetotransport measurements uncover strong interlayer coupling and an anomalous temperature-driven evolution of the Landau-level degeneracy from fourfold to twelvefold, a behavior incompatible with a conventional single-particle description. Together, these results establish $29^\circ$ TBG as a compelling platform for exploring symmetry-driven electronic reconstruction, where quasiperiodicity profoundly reshapes low-energy electronic states beyond the standard moir\'{e} paradigm.\\

\begin{acknowledgments}
We thank L.-Y. Chen, K.-F. Chiu, P. Jarillo-Herrero and M. Koshino for insightful discussion and/or technical support. This research was supported by the National Science and Technology Council in Taiwan (Grant Numbers 114-2123-M-006-001, 114-2112-M-006-029-MY3, 113-2123-M-006-002, 
112-2112-M-006-019-MY3, 112-2628-M-006-001), and the Higher Education Sprout Project, Ministry of Education to the Headquarters of University Advancement at the National Cheng Kung University (NCKU). We also acknowledge National Center for High-performance Computing (NCHC) for providing computational and storage resources. T.-L.C. and M.-W.C. acknowledge support from the National Science and Technology Council and the Ministry of Education in Taiwan. K.W. and T.T. acknowledge support from the JSPS KAKENHI (Grant Numbers 21H05233 and 23H02052), the CREST (JPMJCR24A5), JST and World Premier International Research Center Initiative (WPI), MEXT, Japan.

\end{acknowledgments}

\bibliography{PRL_arXiv_ref}

\nocite{*}

\end{document}